\documentclass[%
longbibliography,reprint,onecolumn,12pt,amsmath,amssymb,aps,]{revtex4-1}

\usepackage{graphicx}% Include figure files
\usepackage{dcolumn}% Align table columns on decimal point
\usepackage{bm}% bold math
\usepackage{color} %%%YS
\usepackage{ulem}

\begin{document}
\newcommand{\sech}{\mathop{\rm sech}\nolimits}
\newcommand{\ba}{\begin{array}}
\newcommand{\ea}{\end{array}}
\newcommand{\be}{\begin{equation}}
\newcommand{\ee}{\end{equation}}
\newcommand{\bea}{\begin{eqnarray}}
\newcommand{\eea}{\end{eqnarray}}
\newcommand{\bfig}{\begin{figure}}
\newcommand{\efig}{\end{figure}}
\newcommand{\Bl}{\Bigl}
\newcommand{\Br}{\Bigr}
\newcommand{\br}{{\bf r}}
\newcommand{\bV}{{\bf V}}
\newcommand{\RE}{{\rm Re}\,}
\newcommand{\IM}{{\rm Im}\,}
\newcommand{\re}{{\rm e}}
\newcommand{\ri}{{\rm i}}
\newcommand{\tF}{\tilde{F}}
\newcommand{\tG}{\tilde{G}}
\newcommand{\tI}{\tilde{I}}
\newcommand{\dd}{{\rm d}}
\newcommand{\al}{\alpha}
\newcommand{\Gm}{\Gamma}
\newcommand{\Lb}{\Lambda}
\newcommand{\vp}{\varphi}
\newcommand{\om}{\omega}
\newcommand{\omh}{\hat{\omega}}
\newcommand{\Om}{\Omega}
\newcommand{\Pih}{\hat{\Pi}}
\newcommand{\din}{\displaystyle\int\limits}
\newcommand{\IN}{\displaystyle\int\limits_{0}^}
\newcommand{\II}{\displaystyle\int\limits^{\infty}_}
\newcommand{\IE}{\displaystyle\int\limits^{1}_}
\preprint{APS/123-QED}
\title{Evolution of non-stationary pulses in a cold magnetized quark-gluon plasma}
\author{D. A. Foga\c{c}a{$^\dag$}, R. Fariello{$^{\ddag\,\dag}$}, F. S. Navarra{$^\dag$} and Y. A. Stepanyants{$^\S$}}
\email{Corresponding author: David Augaitis Foga\c{c}a, e-mail: david@if.usp.br}
\address{{$^\dag$}Instituto de F\'isica, Universidade de S\~ao Paulo,
	Rua do Mat\~ao, 1371,  05508-090 S\~ao Paulo, SP, Brazil}
\address{{$^\ddag$}Departamento de Ci\^encias da Computa\c{c}\~ao, Universidade Estadual de Montes Claros,
	Avenida Rui Braga, sn, Vila Mauriceia, 39401-089, Montes Claros, Minas Gerais, Brazil}
\address{{$^\S$}School of Agricultural, Computational and Environmental Sciences,
	University of Southern Queensland, QLD 4350, Australia and\\
	{$^*$}Department of Applied Mathematics, Nizhny Novgorod
	State Technical University, Nizhny Novgorod, 603950, Russia.}

\begin{abstract}
We study weakly nonlinear wave perturbations propagating in a cold nonrelativistic and magnetized ideal quark-gluon plasma. We show that such perturbations can be described by the Ostrovsky equation. The derivation of this equation is presented for the baryon density perturbations. Then we show that the generalized nonlinear Schr{\"o}dinger (NLS) equation can be derived from the Ostrovsky equation for the description of quasi-harmonic wave trains. This equation is modulationally stable for the wave number $k < k_m$ and unstable for $k > k_m$, where $k_m$ is the wave number where the group velocity has a maximum. We study numerically the dynamics of initial wave packets with the different carrier wave numbers and demonstrate that depending on the initial parameters they can evolve either into the NLS envelope solitons or into dispersive wave trains.
\end{abstract}

%\pacs{PACS Numbers : 12.39.Ba 	(Bag model),
%47.35.Fg (Solitary waves), 12.38.-t(Quantum chromodynamics),
%12.38.Mh(Quark-gluon plasma)  }

\maketitle

\section{Introduction}
\label{Introduction}

\subsection{The plasma of quarks and gluons}

Under usual conditions, all known matter  consists of combinations of atoms. The
atomic nucleus is composed of protons and neutrons, which, in turn, are  made of
quarks. Quarks are pointlike particles and interact with each other through the
color  force. The theory which describes these interactions is called
Quantum Chromodynamics (QCD) and one of its basic assumptions is color
confinement. It implies that particles with color charges, such as quarks and
gluons, can only
exist when combined with other colored particles in such a way that they form  color
neutral objects, which are called hadrons. There are two types of hadrons: those made
by three quarks (as the proton), which are called baryons, and those made by a quark
and an antiquark, which are called mesons.

Quarks and gluons are usually confined in the interior of hadrons,
which have sizes of the order of $1$ Fermi ($= 10^{-15}$ m). Under very special
conditions, QCD predicts that they may form a quark gluon plasma (QGP),
a new state of matter where quarks and gluons are deconfined and are free to travel longer distances.  Presumably, the Universe went through a phase of quark-gluon plasma in the first instants after the Big Bang (from $10^{-11}\, s$ to $10^{-4}\, s$) \cite{florko,aref,we2015}.

In the period $2000-2005$ the QGP was discovered in a series of experiments carried out
at the Brookhaven National Laboratory (USA). There, at the
Relativistic Heavy Ion Collider (RHIC), nuclei were accelerated at very high energies ($200$ $GeV$ per nucleon pair).
The collisions of these nuclei produced an extremely hot medium, whose properties
could only be understood if it were a fluid of deconfined quarks and gluons. In
contrast to the original expectation, these quarks and gluons did not form an ideal
gas of non-interacting particles. Instead, they formed a liquid of deconfined but
strongly interacting particles.

After the discovery a new period started, from $2005$ until today, where the goal
was to determine with more precision the properties of the QGP.  In $2009$ a series of
measurements suggested that there might be waves in this fluid and that we could
observe their effects.  Waves in QGP can be formed when, for example, there is a very
energetic collision between two quarks and after that they have to traverse the fluid.
This exciting possibility prompted a long series of theoretical works on wave
propagation in the QGP, which is a hot topic even nowadays.

In the years $2009-2011$ it was realized that in non head-on heavy ion collisions, when
the impact parameter is small but non-zero, the colliding charged ions produce a very
strong magnetic field. In fact, it can reach the strength of $10^{19}$ Gauss, being
one of the strongest magnetic fields in the Universe.

In this work, motivated by all this accumulated knowledge, we study the propagation of
waves in a quark gluon plasma subjected to a strong magnetic field. This magnetized
QGP can be formed in heavy ion collisions but it can also exist in the interior of
dense and compact stars. In the core of neutron stars baryons are so strongly compressed
that they may  experience a phase transition and form a cold quark gluon plasma.
The existence of a QGP phase in stars is possible,
but not yet experimentally confirmed.   Moreover, it was hypothesised that even quark stars can exist in the Universe \cite{Grimshaw08}.

In the case of the zero magnetic field, it has been shown in Ref. \cite{fnf-11} that weakly nonlinear perturbations in the QGP are described by the Korteweg-de Vries (KdV) equation. This equation admits steadily propagating soliton solutions. This could lead to the remarkable phenomenon: microscopic perturbations propagating through a medium over macroscopic distances. Here we will discuss the influence of a magnetic field on the dynamics of weakly nonlinear perturbations in QGP. It will be shown that in this case the KdV equation should be replaced by such Ostrovsky equation \cite{ostrov} which does not admit soliton solutions \cite{Leonov,Galkin}. Then, we will consider quasi-harmonic wave packets and derive the nonlinear Schr\"{o}dinger (NLS) equation for their description. The solitary waves in this equation can exist in the form of envelope solitons \cite{OstrPotap99,AblSegur81}. We study numerically the Ostrovsky equation with the different initial conditions and demonstrate that such envelope solitons can be formed indeed at certain initial conditions.

\subsection{Korteweg-de Vries solitons in nuclear physics}

As mentioned above, the KdV equation has solitary wave solutions (solitons) which play an important
role in the dynamics of pulse-type initial perturbations. The existence of KdV
solitons in  other fluids made of strongly interacting hadronic matter has been
already investigated. The first works on the subject were published in Refs.
\cite{raha}, where the authors considered the propagation of baryon density pulses
in proton-nucleus collisions at intermediate energies. In this scenario the
incoming proton would be absorbed by the nuclear fluid generating a KdV soliton,
which, traversing the whole nucleus without distortion, would escape from the
target as a proton and would simulate an unexpected transparency. In Ref.
\cite{raha} the existence of KdV solitons relied solely on the equation of state
(EOS), which, however, had no deep justification. In Ref. \cite{fn-06} the authors
have reconsidered the problem, introducing an equation of state derived from the
relativistic mean field models of nuclear matter. It was concluded that the
homogeneous meson field approximation was too strong and would exclude the
existence of KdV solitons. The higher order derivative terms which appear in
the KdV equation could be traced back to the derivative couplings
between the nucleon and vector meson. In Ref. \cite{fn-07} the authors extended the
analysis to the relativistic hydrodynamics and in Ref. \cite{ffn-09} the authors
considered a hadronic matter at the finite temperature and studied the effects of
temperature on the KdV solitons. In Ref. \cite{ffn-10} the authors started to study
perturbations in the QGP at zero and finite temperature. The conclusion of this last
work was that the existence of KdV solitons in a QGP depends on the details of EOS.
In particular, with the EOS derived from the
MIT (Massachusetts Institute of Technology) bag model,
KdV solitons do not exist! A further study of the equation of state carried out in
Ref. \cite{fn-11} showed that if the non-perturbative effects are included in the
EOS through the gluon condensates, then new terms appear in the expressions for the
energy density and pressure and in Ref. \cite{fnf-11} it was shown how these new
terms lead to the KdV equation, after the proper treatment of hydrodynamical equations.

\subsection{The magnetic field and the Ostrovsky equation}

The existence and effects of a magnetic field in quark stars have been studied since long time ago \cite{bfield} and became a hot subject nowadays. In a different context, about ten years ago it was realized that a very strong magnetic field might be produced also in relativistic heavy ion collisions and it might have some effect on the quark-gluon plasma phase \cite{bhic}. An ensuing  question is then: {\it{What is the effect of magnetic field on waves propagating through the QGP}?}

In the recent work \cite{fsn-18}, some of us studied the existence of
stable and causal perturbations in an
ideal, cold, and magnetized quark-gluon plasma.
The dispersion relation for the density and velocity perturbations was derived and it was shown that the existence of a strong magnetic field does not lead to instabilities. Moreover, in  most of the considered cases the propagation of these waves was found to respect causality. A magnetic field changes the pressure, the energy density, and sound speed. It also changes the equations of hydrodynamics. One of the conclusions of Ref. \cite{fsn-18} is that the changes in hydrodynamics are by far more important than the changes in the equation of state.

In the subsequent paper \cite{fsn-19} the study was extended to the case of nonlinear waves. It was investigated the effects of a strong and uniform magnetic field on nonlinear baryon density perturbations in an ideal and magnetized quark-gluon plasma. It was considered the influence of a magnetic field on the EOS and Euler equation. This last study might be applied to the deconfined cold quark matter in compact stars and cold quark-gluon plasma formed in heavy ion collisions at the intermediate energies in the  Facility for Antiproton and Ion Research (FAIR) \cite{fair} or Nuclotron-based Ion Collider fAcility (NICA) \cite{nica}. The reduced Ostrovsky equation (ROE) was derived in this case, which has analytical solutions \cite{Stepanyants, Grimshaw} given by a rarefaction solitonic pulse of the baryon perturbation. In Ref. \cite{fsn-19} a strong mean field approximation was adopted, whereas the effects of density inhomogeneity was neglected.

In the present paper we combine the ideas of Ref. \cite{fnf-11} with the ideas of Refs. \cite{fsn-18} and \cite{fsn-19} and study  nonlinear waves of baryon density perturbation in an inhomogeneous and magnetized QGP. We develop an analytical approach which spreads beyond the mean field approximation for the gluon component \cite{werev} and accounts for a spatial inhomogeneity in the quark matter. The inhomogeneity is described by the Laplacian of baryon density. In the adapted formalism of reductive perturbation method (RPM) \cite{rpm}), the Laplacians yield a new term in the derived wave equation. Essentially, this is the dispersive term proportional to the third spatial derivative of baryon density perturbation, which augments the wave equation previously derived in Ref. \cite{fsn-19}. Then we derive the Ostrovsky equation for the baryon density perturbation in the quark gluon-plasma in the magnetic field.

The Ostrovsky equation was originally derived in the realm of physical oceanography to describe surface and internal waves
in a rotating ocean \cite{ostrov, Ostrov15}. As was shown in Refs. \cite{Leonov, Galkin}, this equation does not possess stationary solitary solutions in the case of ``normal dispersion'', whereas in the case of ``abnormal dispersion'' such solutions can exist \cite{Obregon98, Grimshaw16}. However, as was shown in Ref. \cite{Gilman,OstrovStep16}, solitary waves can exist on long background waves. The Ostrovsky equation and its reduced versions were also derived in various physical contexts, in particular, in relaxing media \cite{Vakhnenko}, plasma physics \cite{Obregon98}, solid state physics (see \cite{Khusnutdinova11} and references therein), elastic layered media \cite{Khusnutdinova17}. It was also shown that this equation is applicable to weakly nonlinear waves of any nature with a small dispersion in random media \cite{BenilovPel}.

A KdV soliton, being not supported by external field, experiences a terminal decay \cite{Grimshaw98, Obregon12, Grimshaw16}. This process ends up by formation of an envelope soliton \cite{rgs, Grimshaw16}; this was confirmed experimentally in Ref. \cite{Grimshaw13}. We will follow these works in searching for the solutions of Ostrovsky equation derived here for the QGP.

In the next section we present the basic nonrelativistic hydrodynamic equations in the presence of an external magnetic field. In Section III we introduce the equation of state with the external magnetic field. In  Section IV we derive the Ostrovsky equation via RPM and in Section V we follow the approach developed in Refs. \cite{rgs} to obtain numerical solutions of the Ostrovsky equation. Finally we present some comments and conclusions.

\section{Nonrelativistic hydrodynamics with an external magnetic field}
\label{Sect1}

We start with the nonrelativistic Euler equation \cite{land} with an external uniform magnetic field \cite{we16,fsn-18,fsn-19}.  The  magnetic field is included also in the equation of state. The uniform magnetic field has intensity $B$ and directed along the $z$-axis: $\vec{B}=B \hat{z}$. As usual, we consider the three quarks species for the quark-gluon plasma (QGP): up ($u$), down ($d$), and strange ($s$) quarks with the charges $Q_{u}= 2 \, Q_{e}/3$, $Q_{d}= - \, Q_{e}/3$  and $Q_{s}= - \, Q_{e}/3$ respectively, with the corresponding masses: $m_{u}=2.2 \, MeV$, $m_{d}=4.7 \, MeV$, and $m_{s}=96 \, MeV$ \cite{pdg}. Throughout this work, we employ natural units ($\hbar=c=1$); the metric used is $g^{\mu\nu}=\textrm{diag}(+,-,-,-)$, and the absolute value of electron charge is $Q_{e}=0.08542$ \cite{glend}. The quark mass density $({\rho_{m}}_f)$ and the baryon density $({\rho_{B}}_{f})$ are related by the equation ${\rho_{m}}_f=3\,m_{f} \, {\rho_{B}}_{f}$. The charge density for each quark is ${\rho_{c}}_{f}=3\,{Q_{f}} \, {\rho_{B}}_{f}$.

Due to the influence of an external magnetic field, the quarks with different charges move along different trajectories, therefore the multi-fluid approach should be used \cite{azam,multif,we16,fsn-18,fsn-19}. The Euler equation for each quark flavour $f$ with an external uniform magnetic field has the form \cite{land,fsn-18, fsn-19}:
\begin{equation}
	3\,m_{f} \, {\rho_{B}}_{f}\Bigg[{\frac{\partial \vec{v_f}}{\partial t}} +
	(\vec{v_f}\cdot \vec{\nabla}) \vec{v_f}\Bigg]=
	-\vec{\nabla}p
	+3\,{Q_{f}} \, {\rho_{B}}_{f}\vec{v_f} \times \vec{B}.
	\label{eulermag}
\end{equation}

The pressure gradient in the presence of magnetic field is anisotropic \cite{fsn-19}:
\begin{equation}
	\vec{\nabla}{p}=\Bigg({\frac{\partial}{\partial x}}\,
	{p_{\perp}}\,,\,{\frac{\partial }{\partial y}}\,{p_{\perp}}\,,\,
	{\frac{\partial }{\partial z}}\,{p_{\parallel}}\Bigg),
	\label{gradepresses}
\end{equation}
where $p_{f\,\parallel}$ is the pressure in the direction of magnetic field, and $p_{f\,\perp}$ is the  pressure perpendicular to the magnetic field.

The continuity equation for the baryon density reads \cite{land,fsn-19,fsn-18}:
\begin{equation}
	{\frac{\partial {\rho_{B}}_{f}}{\partial t}} + \vec{\nabla}
	\cdot ({\rho_{B}}_{f} \,{\vec{v_f}})=0.
	\label{contmag}
\end{equation}
To simplify the notations, from now on we omit the flavour label ``f'', and later we will specify it when necessary.

\section{The equation of state}
\label{Sect2}

The first derivation of equation of state, which was dubbed the modified quantum chromodynamics (mQCD), was presented in \cite{fn-11}. The mQCD with an
external magnetic field was deduced in \cite{we16} and used in the calculation of stellar structure of compact quark stars under the magnetic field effect \cite{we16}. It was applied to study of causality and stability of waves \cite{fsn-18} and more recently, to study the nonlinear effects in baryon density waves in the mean field theory (MFT) approach \cite{fsn-19}.

Here we repeat similar calculations which were previously performed in Refs. \cite{we16,fn-11}, but with a different approach for the gluon field ${\alpha}^{a}_{0}$. In contrast to earlier assumed spatially homogeneous gluon field, now we consider that it can be a function of space and time, ${\alpha}^{a}_{0}(x,y,z,t)$. Going beyond
the usual MFT approach, we are able to include the inhomogeneities of the baryon density
in the energy density and in the pressure. This is done as follows. We start from
the equation of motion of the (massive) gluon field:
\begin{equation}
	-{\vec{\nabla}}^{2}{\alpha}^{a}_{0}+{m_{G}}^{2}{\alpha}^{a}_{0}=-g \rho^{a},
	\label{azeroeq}
\end{equation}
where $m_G$ and $g$ are the gluon mass and the quark-gluon coupling constant respectively.

Assuming that the spatial variation of the gluon field is very smooth and neglecting the derivatives, we find in the zero approximation the mean gluon field:
\begin{equation}
	{\alpha}^{a}_{0}\cong -{\frac{g}{{m_{G}}^{2}}}\rho^{a}.
	\label{azerofirst}
\end{equation}

In the next approximation we insert solution (\ref{azerofirst}) into the term ${\vec{\nabla^{2}}}{\alpha}^{a}_{0}$ of Eq. (\ref{azeroeq}) and solve it again for the gluon field ${\alpha}^{a}_{0}$ \cite{werev}:
\begin{equation}
	{\alpha}^{a}_{0}=-{\frac{g}{{m_{G}}^{2}}}\rho^{a}
	-{\frac{g}{{m_{G}}^{4}}}{\vec{\nabla^{2}}}\rho^{a}.
	\label{azerofinal}
\end{equation}
Here $\rho^{a}$ is the temporal component of the color vector current $j^{a\nu}$, given by $j^{a0}=\rho^{a}=\sum_{f}{\bar{\psi}_{i}}^{f} \gamma^{0}T^{a}_{ij}\psi_{j}^{f} =\sum_{f}{{\psi}_{i}^{\dagger}}^{f}T^{a}_{ij}\psi_{j}^{f}$ and related to total net quark density $\rho$ by the equation $\rho^{a}\rho^{a}=3{\rho}^{2}/8$. The total net quark density in turn is related with the baryon density $\rho_{B}$ through the following relation $\rho_{B}=\rho/3$, see the Appendix in Ref. \cite{we16}.

After some algebra (the details can be found in Refs. \cite{we16,fn-11}), the energy density ($\varepsilon$), parallel pressure ($p_{f\,\parallel}$), and perpendicular pressure ($p_{f\,\perp}$) can be presented respectively as:
$$
\varepsilon = {\frac{27{g}^{2}}{16{m_{G}}^{2}}}\,\left[({{\rho_{B}}})^{2}
+ {\frac{\rho_{B}}{{m_{G}}^{2}}}\,{\vec{\nabla^{2}}}
{{\rho_{B}}} + {\frac{\rho_{B}}{{m_{G}}^{4}}}\,{\vec{\nabla^{2}}}\,
\Big({\vec{\nabla^{2}}}{{\rho_{B}}}\Big) + {\frac{1}{{m_{G}}^{6}}}\Big({\vec{\nabla^{2}}}{{\rho_{B}}}\Big)
\,{\vec{\nabla^{2}}}\,\Big({\vec{\nabla^{2}}}\rho_{B}\Big)\right]
$$
\begin{equation}
{} + {\mathcal{B}}_{QCD}+{\frac{B^{2}}{8\pi}}
	+\sum_{f=u}^{d,s}{\frac{|Q_{f}|B}{2\pi^{2}}} \, \sum_{n=0}^{n^{f}_{max}}
	3(2-\delta_{n0}) \int_{0}^{k^{f}_{z,F}} dk_{z}\sqrt{m_{f}^{2}+k_{z}^{2}+2n|Q_{f}|B},
	\label{eps}
\end{equation}
\\$$
p_{\parallel} = {\frac{27{g}^{2}}{16{m_{G}}^{2}}}\,\left[({{\rho_{B}}})^{2}
+ \frac{\rho_{B}}{{m_{G}}^{2}}\,{\vec{\nabla^{2}}}{\rho_{B}}
- {\frac{\rho_{B}}{{m_{G}}^{4}}}\,{\vec{\nabla^{2}}}\,\Big({\vec{\nabla^{2}}}{\rho_{B}}\Big)
- {\frac{1}{{m_{G}}^{6}}}\Big({\vec{\nabla^{2}}}{\rho_{B}}\Big)
\,{\vec{\nabla^{2}}}\,\Big({\vec{\nabla^{2}}}{\rho_{B}}\Big)\right]
$$
\begin{equation}
{} - {\mathcal{B}}_{QCD}-{\frac{B^{2}}{8\pi}}
	+\sum_{f=u}^{d,s}{\frac{|Q_{f}|B}{2\pi^{2}}} \, \sum_{n=0}^{n^{f}_{max}}
	3(2-\delta_{n0}) \int_{0}^{k^{f}_{z,F}} dk_{z} \, {\frac{{k_{z}}^{2}}
		{\sqrt{m_{f}^{2}+k_{z}^{2} + 2n|Q_{f}|B}}},
\label{parap}
\end{equation}
\\
$$
p_{\perp} = {\frac{27{g}^{2}}{16{m_{G}}^{2}}}\,\left[({{\rho_{B}}})^{2}
+ \frac{\rho_{B}}{{m_{G}}^{2}}\,{\vec{\nabla^{2}}}{\rho_{B}}
- \frac{\rho_{B}}{{m_{G}}^{4}}\,{\vec{\nabla^{2}}}\,
\Big({\vec{\nabla^{2}}}{\rho_{B}}\Big) - \frac{1}{{m_{G}}^{6}}\Big({\vec{\nabla^{2}}}{\rho_{B}}\Big)
\,{\vec{\nabla^{2}}}\,\Big({\vec{\nabla^{2}}}{\rho_{B}}\Big)\right]
$$
\begin{equation}
{} - {\mathcal{B}}_{QCD}+{\frac{B^{2}}{8\pi}}
	+\sum_{f=u}^{d,s}{\frac{|Q_{f}|^{2}B^{2}}{2\pi^{2}}} \, \sum_{n=0}^{n^{f}_{max}}
	3(2-\delta_{n0}) n \int_{0}^{k^{f}_{z,F}} \, {\frac{dk_{z}}{\sqrt{m_{f}^{2} + k_{z}^{2}+2n|Q_{f}|B}}}.
\label{perp}
\end{equation}
In these equations the terms ${\vec{\nabla^{2}}}{\rho_{B}}$ came from the formula ${\vec{\nabla^{2}}}\rho^{a}$ in Eq. (\ref{azerofinal}). Again, as in Ref. \cite{fsn-19}, the baryon density is:
\begin{equation}
	\rho_{B}=\sum_{f=u}^{d,s}\,{\frac{|Q_{f}|B}{2\pi^{2}}} \,
	\sum_{n=0}^{n^{f}_{max}}(2-\delta_{n0}) \, \sqrt{{\nu_{f}}^{2}-m_{f}^{2}-2n|Q_{f}|B}
	\label{magbard}
\end{equation}
$$
\mbox{with} \quad n\leq n^{f}_{max} = \Bigg\lfloor\frac{{\nu_{f}}^{2}-m_{f}^{2}}{2|Q_{f}|B}\Bigg\rfloor,
$$
where the symbol $\lfloor\ldots\rfloor$ stands for the integer part of the corresponding expression $(\ldots)$, and ${\nu_{f}}$  is the chemical potential for the quark of the flavour $f$. As usual, for a
fixed magnetic field intensity, we choose the chemical potential ${\nu_{f}}$ and
then determine $\rho_{B}$. Now we need to fix the value of the coupling constant $g$, the dynamical gluon mass $m_{G}$, and also the ``bag'' constant ${\mathcal{B}}_{QCD}$. When $g/m_{G} \rightarrow 0$ we recover the equation of state for the MIT model \cite{fsn-18,we16}.

\section{The Ostrovsky equation and nonlinear Schr\"{o}dinger equation}
\label{Sect3}

As shown in Appendix, weakly nonlinear perturbations of any quark flavour $f$ in QGP are described by the Ostrovsky equation:
\begin{equation}
{\frac{\partial}{\partial \xi}}\Bigg({\frac{\partial\eta}{\partial t}} + c_{0}\,{\frac{\partial\eta}{\partial \xi}} + \alpha\,\eta{\frac{\partial\eta}{\partial \xi}} + \beta\,{\frac{\partial^{3}\eta}{\partial \xi^{3}}}\Bigg) = \Gamma\,\eta,
	\label{ostrovsky}
\end{equation}
where $\eta \equiv {\delta\rho_{B}}_{f}$ (for simplicity in the notation, we omit the index $f$ in all positive coefficients $c_{0}$, $\alpha$, $\beta$, and $\Gamma$ in this equation. The expressions for the coefficients are:
\begin{equation}
	c_{0} = 2{c_{s\,\perp}} \,\, , \,\,\,\,
	\alpha = {\frac{3}{2}}{c_{s\,\perp}} \,\, , \,\,\,\,
	\beta = {\frac{{c_{s\,\perp}}}{4\,{m_{G}}^{2}}},
	\,\,\,\,\,\, \textrm{and} \,\,\,\,\,\,\,
	\Gamma = \frac{(Q_f \,B)^2 }{2 \, m_f^2 \,c_{s\,\perp}},
	\label{ostrocoefs}
\end{equation}
where $c_{0}$ is the velocity of dispersionless linear waves, $\alpha$ is the nonlinear coefficient, $\beta$ is the coefficient of a small-scale dispersion, and $\Gamma$ is the coefficient of large-scale dispersion.

When the derivative terms ${\vec{\nabla^{2}}}{\rho_{B}}$ in Eqs. $(\ref{parap})$ and $(\ref{perp})$ are neglected, then we recover the reduced Ostrovsky equation (ROE) \cite{fsn-19} with $\beta=0$ in Eq. (\ref{ostrovsky}). When there is no magnetic field, the $\Gamma=0$, and Eq. (\ref{ostrovsky}) reduces to the classical KdV equation \cite{drazin}.

As has been mentioned in the Introduction, when the coefficients of the Ostrovsky equation (\ref{ostrovsky}) is such that $\beta\Gamma>0$, then the ``antisoliton theorem'' is fulfilled \cite{Leonov, Galkin}, which states that there are no stationary soliton solutions. If the right-hand side in Eq. (\ref{ostrovsky}) is small in comparison with the nonlinear and dispersive term in the left-hand side, then the initial KdV soliton experiences a terminal decay \cite{Grimshaw98, Obregon12, Grimshaw16} and formally disappears in finite time. However, the dynamics of a wave perturbation after the extinction time is more complicated (see, for example, Ref. \cite{OstrStep90, Gilman}) and eventually ends up with the emergence of an envelope soliton described by the nonlinear Schr\"{o}dinger (NLS) equation or its generalisation \cite{rgs, Whitfield, Grimshaw16}.

Originally it was assumed that the envelope soliton appears with the wave number corresponding to the maximum of group velocity of linear waves \cite{rgs}. At such wave number the dispersion coefficient in the classical NLS equation vanishes, therefore the authors of Ref. \cite{rgs} derived the generalized third-order NLS equation and obtained its soliton solution.  However, later a more accurate analysis based on the numerical solutions \cite{ Whitfield} revealed that the wave number of the carrier wave in the envelope is slightly shifted to the higher values where the classical NLS equation is applicable. The shifted value of the wave number was not predicted, but empirically taken from the numerical results. In Ref. \cite{Stepan19} it was assumed that an envelope soliton emerges from a quasi-linear wave field with the wave number corresponding to the maximum of the growth rate of modulation instability. Below we will present the wave number of carrier wave in terms of coefficients of the Ostrovsky equation (\ref{ostrovsky}). Then, we will present the results of numerical study of wave packet evolution within the Ostrovsky equation. Before that, we will discuss the dispersion properties of small-amplitude linear waves and present the NLS equation for narrow-band quasi-harmonic perturbations.

\subsection{Dispersion properties of Ostrovsky equation}
\label{Subsect3.1}

For the wave perturbations of infinitesimal amplitude one can neglect the nonlinear term $\sim\alpha$ in Eq. (\ref{ostrovsky}) and seek for a solution in the form $\eta = A e^{i(k\,\xi-\omega t)}$, where $A$ is the amplitude, $\omega$ is the frequency and $k$ is the wave number. Then we obtain the dispersion relation:
\begin{equation}
	\omega(k)=c_{0}\,k-\beta\, k^{3}+{\frac{\Gamma}{k}}.
	\label{disprela}
\end{equation}

From this equation we derive the phase velocity $V_{p}$ and group velocity $V_{g}$ which are given respectively by the following formulae:
\begin{eqnarray}
V_{p} &=& {\frac{\omega}{k}}=c_{0}-\beta\, k^{2}+{\frac{\Gamma}{k^{2}}}, \label{phasev} \\
%&& \nonumber \\
V_{g} &=& {\frac{\partial \omega}{\partial k}}=c_{0}-3\beta\,k^{2}-{\frac{\Gamma}{k^{2}}}. \label{groupv}
\end{eqnarray}

The group velocity has a maximum at $k_m = (\Gamma/3\beta)^{1/4}$, so that $V'_{g}(k_m) = 0$.

\subsection{The nonlinear Schr\"{o}dinger equation}
\label{Subsect3.2}

Consider now a quasi-monochromatic perturbation with slowly varying amplitude $\eta(\xi, t) = A(\xi) e^{i(k\,\xi-\omega t)}$. As has been shown in Ref. \cite{rgs}, such wave train with the wave number $k \approx k_m$ can be described by the following third-order NLS equation:
\begin{equation}
i\left({\frac{\partial A}{\partial t}} + {V_{g}}\,{\frac{\partial A}{\partial \xi}}\right) + p{\frac{\partial^{2} A}{\partial \xi^{2}}} - iq{\frac{\partial^{3} A}{\partial \xi^{3}}} + \mu_0\,|A|^{2} A + i\left(\mu_{1}|A|^{2} {\frac{\partial A}{\partial \xi}}
	+\mu_{2} A^{2} {\frac{\partial A^{*}}{\partial \xi}}\right) = 0,
	\label{nls}
\end{equation}
where the symbol star stands for complex conjugate, the dispersion coefficients are:
$$
p = \frac{1}{2}\frac{\partial^2 \omega}{\partial k^2} = -3\beta k + \frac{\Gamma}{k^3}, \quad q = \frac{1}{6}\frac{\partial^3 \omega}{\partial k^3} = -\beta - \frac{\Gamma}{k^4},
$$
and the nonlinear coefficients are:
\begin{equation}
\mu_0 = -\frac{2}{3}\frac{\alpha^{2}k^3}{\Gamma + 4\beta k^4}, \quad
\mu_{1} = \frac{2}{3}\frac{\alpha^{2}k^2\left(5\Gamma + 4\beta k^4\right)}{\left(\Gamma + 4\beta k^4\right)^2}, \quad \textrm{and} \quad \mu_{2} = \frac{2}{3}\frac{\alpha^{2}k^2}{\Gamma + 4\beta k^4}.	 %
\label{nlscoefs}
\end{equation}

When $k = k_m$, the coefficients of Eq. (\ref{nls}) simplify and reduce to:
\begin{equation}
p = 0, \quad q = -4\beta, \quad \mu_0 = \frac{-2\alpha^2\sqrt{\Gamma}}{7(3\beta \Gamma)^{3/4}}, \quad
\mu_{1} = {\frac{38\alpha^{2}} {49(3\beta \Gamma)^{1/2}}}, \quad \textrm{and} \quad
\mu_{2} = {\frac{2\alpha^{2}}{7(3\beta \Gamma)^{1/2}}}.	%
\label{nlscoefs1}
\end{equation}

However, if the wave number of a carrier wave is not too close to $k_m$, then the higher-order terms $\sim q$, $\mu_1$, and $\mu_2$ can be neglected, and Eq. (\ref{nls}) reduces to the conventional NLS equation:
\begin{equation}
i\left({\frac{\partial A}{\partial t}} + {V_{g}}\,{\frac{\partial A}{\partial \xi}}\right) + p(k){\frac{\partial^{2} A}{\partial \xi^{2}}} + \mu_0(k)\,|A|^{2} A = 0.
\label{StandardNLS}
\end{equation}

According to the Lighthill criterion \cite{Lighthill, OstrPotap99}, a quasi-monochromatic wave is unstable with respect to self-modulation if $p(k)\mu_0(k) > 0$. In our case this gives:
\begin{equation}
\frac{2\alpha^2 \beta k^4}{\Gamma + 4\beta k^4}\left(1 - \frac{k_m^4}{k^4}\right) > 0.
\label{Lighthill}
\end{equation}
This inequality is true for wave numbers $k > k_m$. The growth rate of modulation instability is:
\begin{equation}
\lambda = \sqrt{2p(k)K^2\left[\mu_0(k) A_0^2 - p(k)K^2/2\right]},
\label{Growth}
\end{equation}
where $A_0$ is the amplitude of a uniform sinusoidal wave train, and $K$ is the wave number of a modulation. The most unstable modulation corresponds to $K^2_{max} = \mu(k) A_0^2/p(k)$, and the maximum of growth rate as the function of a wave number is:
\begin{equation}
\lambda_{max}(k) = -\mu(k) A_0^2 = \frac{2}{3}\frac{\alpha^2 A_0^2 k^3}{\Gamma + 4\beta k^4}.
\label{GrowthMax}
\end{equation}

This function can be further optimized with respect to the wave number of carrier wave $k$; its maximum occurs at $k = k_c \equiv (3\Gamma/4\beta)^{1/4}$ where
\begin{equation}
\lambda_{max}(k_c) = \frac{\alpha^2 A_0^2}{6\Gamma}\left(\frac{3\Gamma}{4\beta}\right)^{3/4}.
\label{GrowthMax1}
\end{equation}

The ratio of critical wave numbers $k_c/k_m = \sqrt{3/2} \approx 1.22$, and the relative difference of wave numbers is:
\begin{equation}
\frac{k_c - k_m}{k_m} = \sqrt{3/2} - 1 \approx 0.22.
\label{ShiftK}
\end{equation}
This difference is quite notable, therefore the wave train with the carrier wave number $k = k_c$ should be described rather by the conventional NLS equation (\ref{StandardNLS}) than by the third-order NLS equation (\ref{nls}).

\subsection{Soliton solutions to the conventional and generalized nonlinear Schr\"{o}dinger equations}
\label{Subsect3.3}

In this subsection we present soliton solutions to the conventional NLS equation (\ref{StandardNLS}) and its generalized version, the third-order NLS equation (\ref{nls}). Solution of Eq. (\ref{StandardNLS}) in the form of envelope soliton is very well known (see, e.g., \cite{AblSegur81, OstrPotap99}):
\begin{equation}
A = A_0\sech{\frac{\xi - Vt}{\Delta}}\exp{\left\{i\left[\left( k + \kappa\right)\xi - \left(\omega + \sigma\right) t\right]\right\}},
\label{EnvNLSSol}
\end{equation}
where half-width of the envelope pulse $\Delta$, gauge $\kappa$, and chirp $\sigma$ are determined by the soliton amplitude $A_0$ and velocity $V$:
\begin{equation}
\Delta = \frac{1}{A_0}\sqrt{\frac{2p}{\mu_0}}, \quad \kappa = \frac{V - V_g}{2p}, \quad \sigma = \frac{V^2 - V_g^2}{4p} - \frac{\mu_0 A_0^2}{2}
\label{NLSPar}
\end{equation}

The soliton solution to the third-order NLS equation (\ref{nls}) was obtained in Ref. \cite{rgs}; it has the same form as Eq. (\ref{EnvNLSSol}), but different relationship between the parameters (there are several typos in Ref. \cite{rgs} which are corrected below):
$$
\Delta = \frac{1}{A_0}\sqrt{\frac{-6q}{\mu_1 + \mu_2}}, \quad \kappa = -\frac{3\mu_0 q + p\left(\mu_1 + \mu_2\right)}{6q\mu_2},
$$
\begin{equation}
\sigma = \frac{\mu_1 + \mu_2}{6q}A_0^2\left(p + 3\kappa q\right) + \kappa\left(V_g + p\kappa + q\kappa^2\right), \quad V = V_g + \frac{\mu_1 + \mu_2}{6}A_0^2 + \kappa\left(2p  + 3\kappa q\right).
\label{TNLSPar}
\end{equation}
Note that this solution contains only one free parameter, for example, soliton amplitude $A_0$, whereas other parameters can be defined in therms of $A_0$, except the gauge $\kappa$ which is determined solely by the coefficients of third-order NLS equation (\ref{nls}).

In the particular case when $p = 0$, i.e. when $k = k_m$, these dependencies simplify and reduce to:
$$
\Delta = \frac{1}{A_0}\sqrt{\frac{-6q}{\mu_1 + \mu_2}}, \quad \kappa = \frac{-\mu_0}{2\mu_2}, \quad \sigma = \frac{\kappa}{2} A_0^2\left(\mu_1 + \mu_2\right) + \kappa\left(V_g + q\kappa^2\right),
$$
\begin{equation}
V = V_g + \frac{\mu_1 + \mu_2}{6}A_0^2 + 3\kappa^2 q.
\label{TNLSPar1}
\end{equation}

This solution, apparently, can exist if its amplitude Fourier spectrum centered at $k = k_m + \kappa$ is narrow enough and vanishes when $k \to k_{m+}$. Otherwise, if the spectrum is sufficiently wide and contains components with $k < k_m$, i.e. beyond the range of self-modulation, then such solution hardly can be stable with respect to small perturbations. Therefore, soliton solution (\ref{EnvNLSSol}), (\ref{TNLSPar1}) with the relatively wide spectrum is, apparently, marginally stable; our numerical study confirms this hypothesis (see below). In Section \ref{Sect4} we present the results of numerical study of evolution of initial pulses with the different carrier waves and compare the outcomes.

\section{Numerical study}
\label{Sect4}

\begin{figure}[h!]
	{\centerline{\includegraphics [scale=0.75]{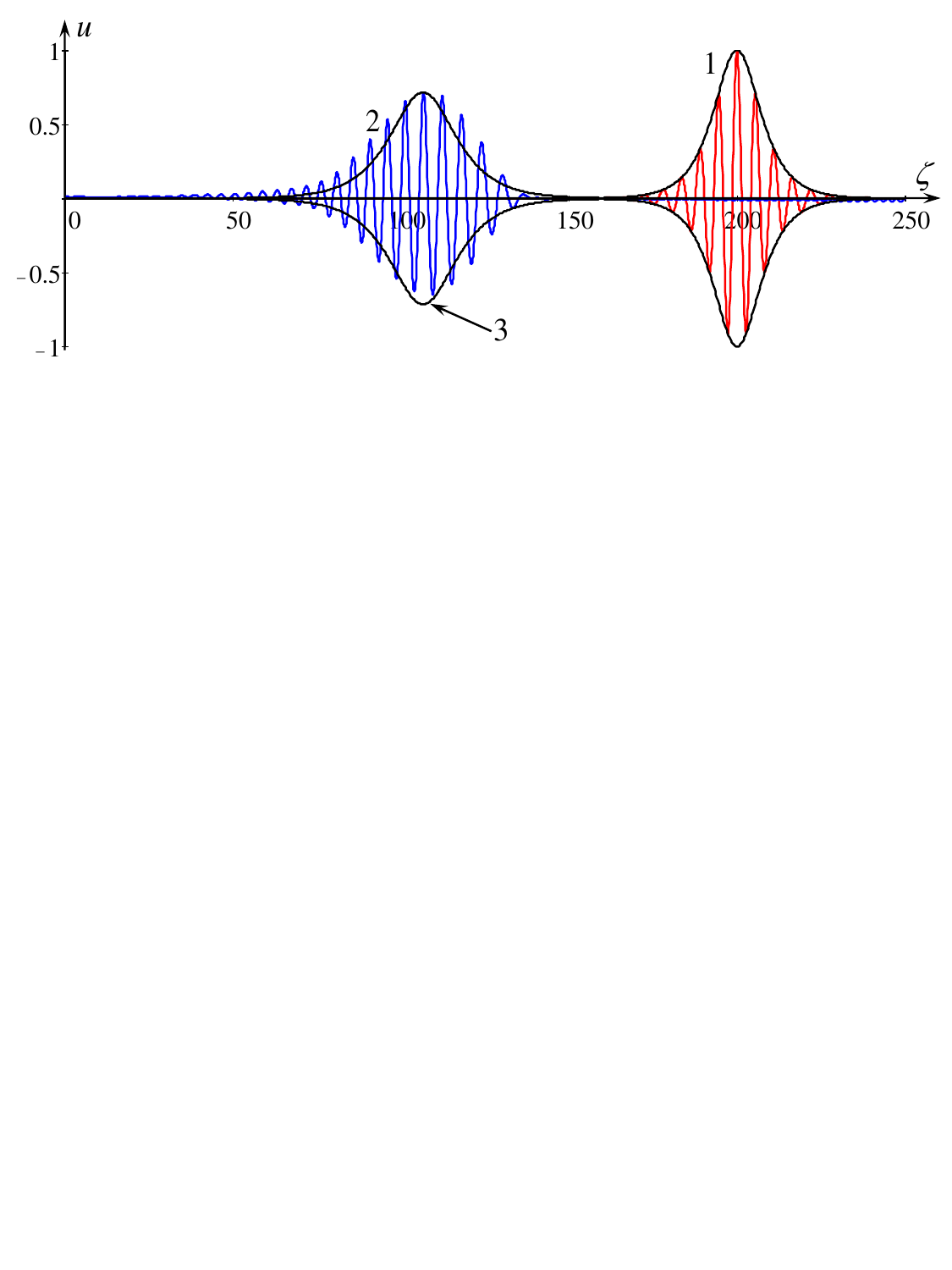}}} %
	\vspace{-13.cm}%
	\caption{\small (Color online) The initial pulse (line 1) in the form of envelope soliton (\ref{EnvNLSSol}) with the parameters given by Eq. (\ref{TNLSPar1}) and amplitude $A_0 = 1$. Line 2 shows the wave train evolving from the initial pulse 1 at $\tau = 10$. Line 3 shows the envelope of a soliton (\ref{EnvNLSSol}) with the same amplitude as the wave train 2. Only a fragment of the total spatial interval $L = 400$ is shown.} %
	\label{f01}
\end{figure}

\begin{figure}[h!]
	{\centerline{\includegraphics [scale=0.75]{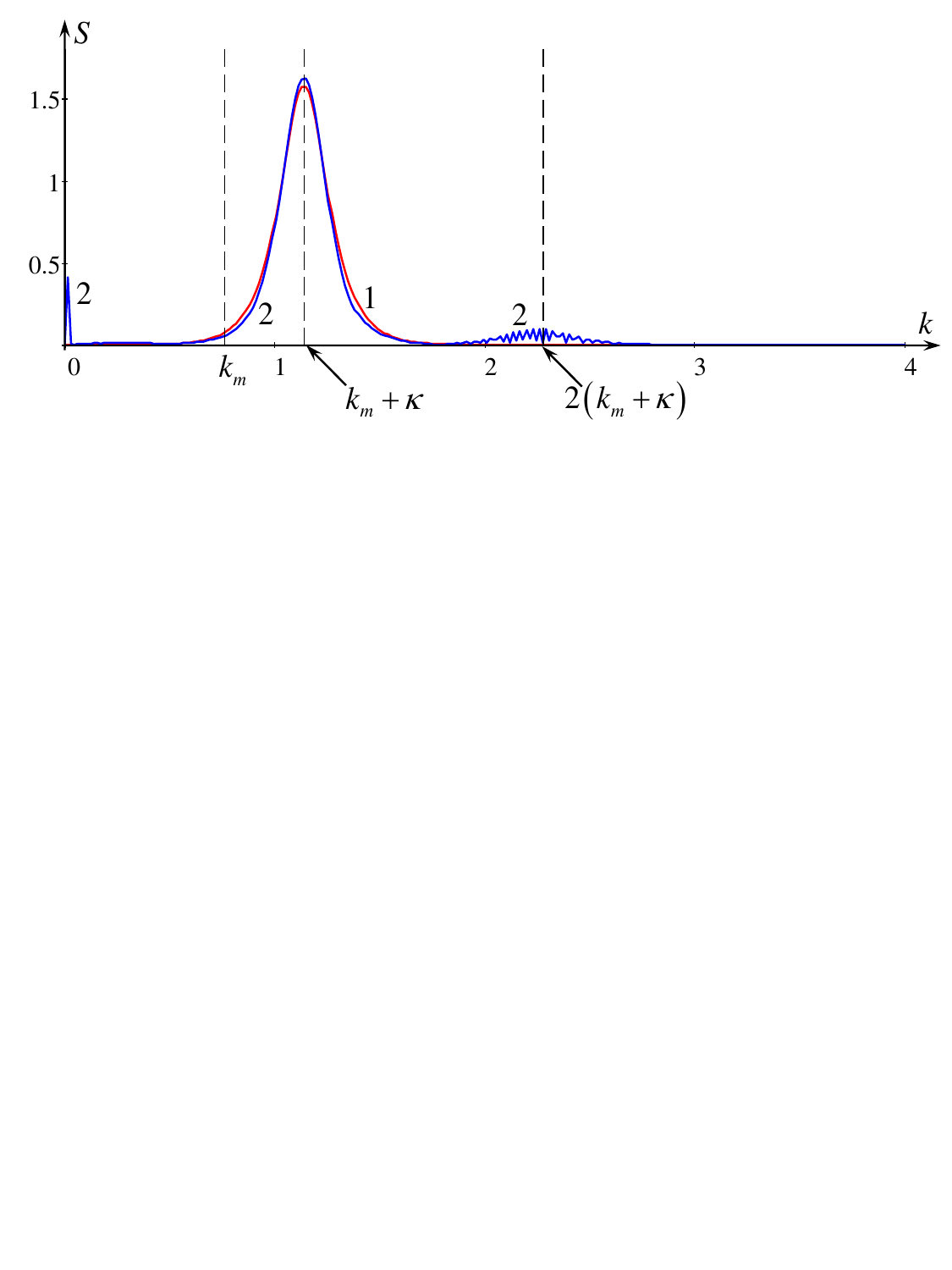}}} %
	\vspace{-13.cm}%
	\caption{\small (Color online) The amplitude Fourier spectrum of soliton solution (\ref{EnvNLSSol}), (\ref{TNLSPar1}) at the initial instant of time (line 1) and spectrum of the pulse shown in Fig. \ref{f01} at $\tau = 10$ (line 2).} %
	\label{f02}
\end{figure}

In this section we present the results of direct numerical modeling of Ostrovsky equation (\ref{ostrovsky}). As the first step, we rewrite this equation in the dimensionless form:
\begin{equation}
\frac{\partial}{\partial \zeta}\Bigg({\frac{\partial u}{\partial \tau}} + u\frac{\partial u}{\partial \zeta} + \frac{\partial^{3} u}{\partial \zeta^{3}}\Bigg) = u,
\label{DLOstrEq}
\end{equation}
where $\tau = \Gamma t (\beta/\Gamma)^{1/4}$, $\zeta = (\Gamma/\beta)^{1/4}\left(\xi - c_0 t\right)$, $u = \alpha\eta/\sqrt{\beta\Gamma}$. The Ostrovsky equation (\ref{DLOstrEq}) was numerically solved by means of the finite-difference  scheme described in Ref. \cite{Obregon12}. To check our hypothesis about the stability of envelope soliton (\ref{EnvNLSSol}), (\ref{TNLSPar1}) with $k = k_m + \kappa$, we use this solution with $A_0 = 1$, $k = k_m = 1/\sqrt[4]{3}$, and $\kappa = k_m/2$ as the initial condition for Eq. (\ref{DLOstrEq}) on the interval $L = 400$ with the periodic boundary conditions (see pulse 1 in Fig. \ref{f01}). The amplitude Fourier spectrum of soliton solution with such parameters is shown in Fig. \ref{f02} by line 1.

\begin{figure}[h!]
	{\centerline{\includegraphics [scale=0.75]{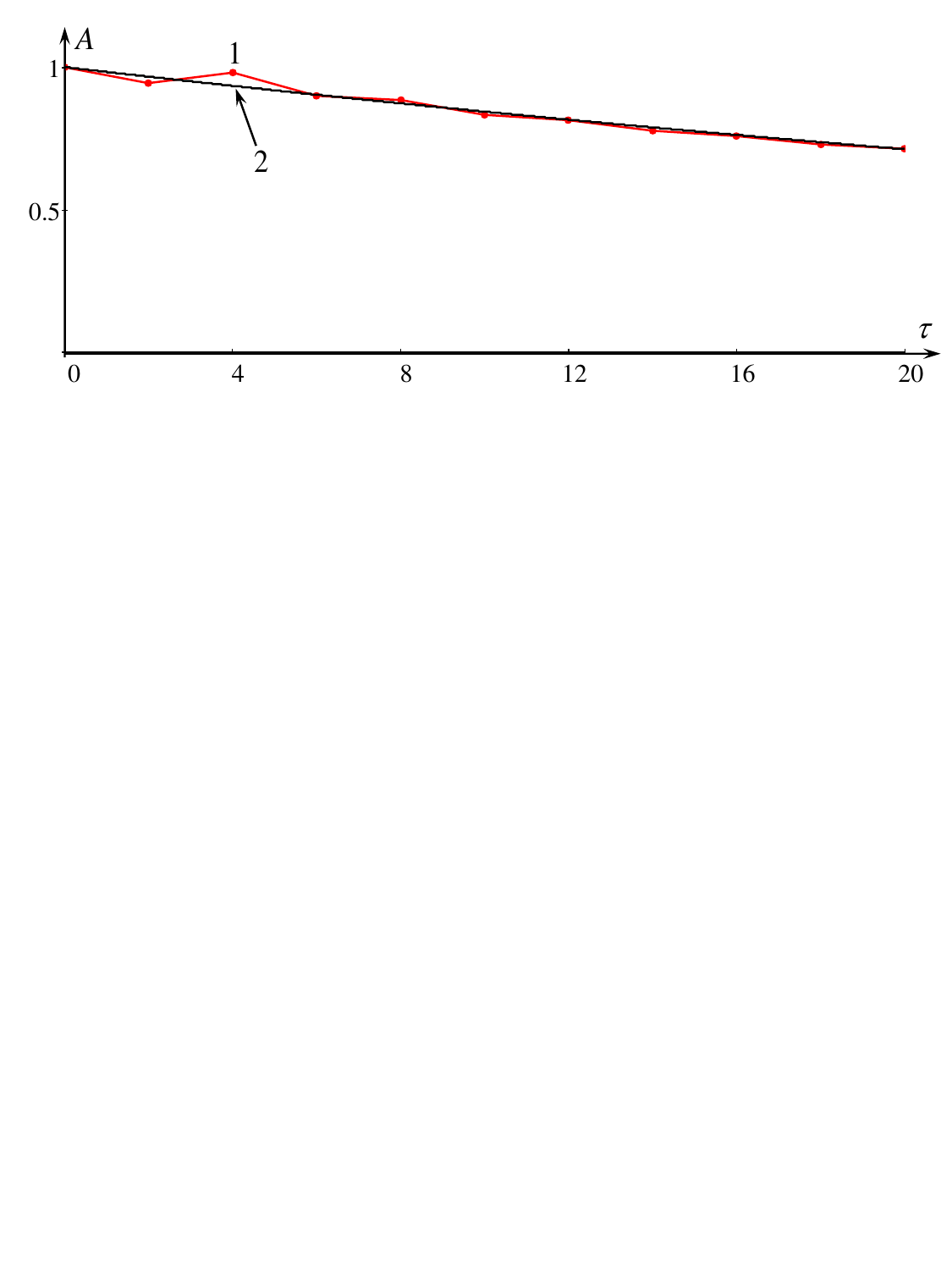}}} %
	\vspace{-13.cm}%
	\caption{\small (Color online) The dependence of wave train amplitude $A(\tau)$ on time in the process of its decay (line 1). Line 2 shows the best fit approximation $A(\tau) = A_0e^{-0.017\tau}$.} %
	\label{f03}
\end{figure}

It is clearly seen that some portion of the initial spectrum on the left from the dashed vertical line 1 where $k < k_m$ is in the range where the modulation instability does not occur. Because of that, the envelope soliton (\ref{EnvNLSSol}) does not remain stationary and undergoes gradual decay. In Fig. \ref{f01} line 2 shows the wave train evolving from the initial envelope soliton by $\tau = 10$. The shape of this wave train does not correlate with the envelope of a soliton of the same amplitude shown by line 3 in the figure. In the process of pulse evolution its spectrum becomes slightly narrower and taller (see line 2 in Fig. \ref{f02}); this can be explained by the manifestation of self-modulation which should lead to the formation of genuine envelope soliton of a narrow spectrum centered at $k = k_m + \kappa$. Due to the influence of nonlinearity we observe also a generation of the second harmonics and sub-harmonics in the vicinity of $k = 0$ (see line 2). Figure \ref{f03} shows pulse amplitude decay with time (dotted line 1) and best fit approximation $A(\tau) = A_0e^{-0.017\tau}$.

\begin{figure}[h!]
	{\centerline{\includegraphics [scale=0.75]{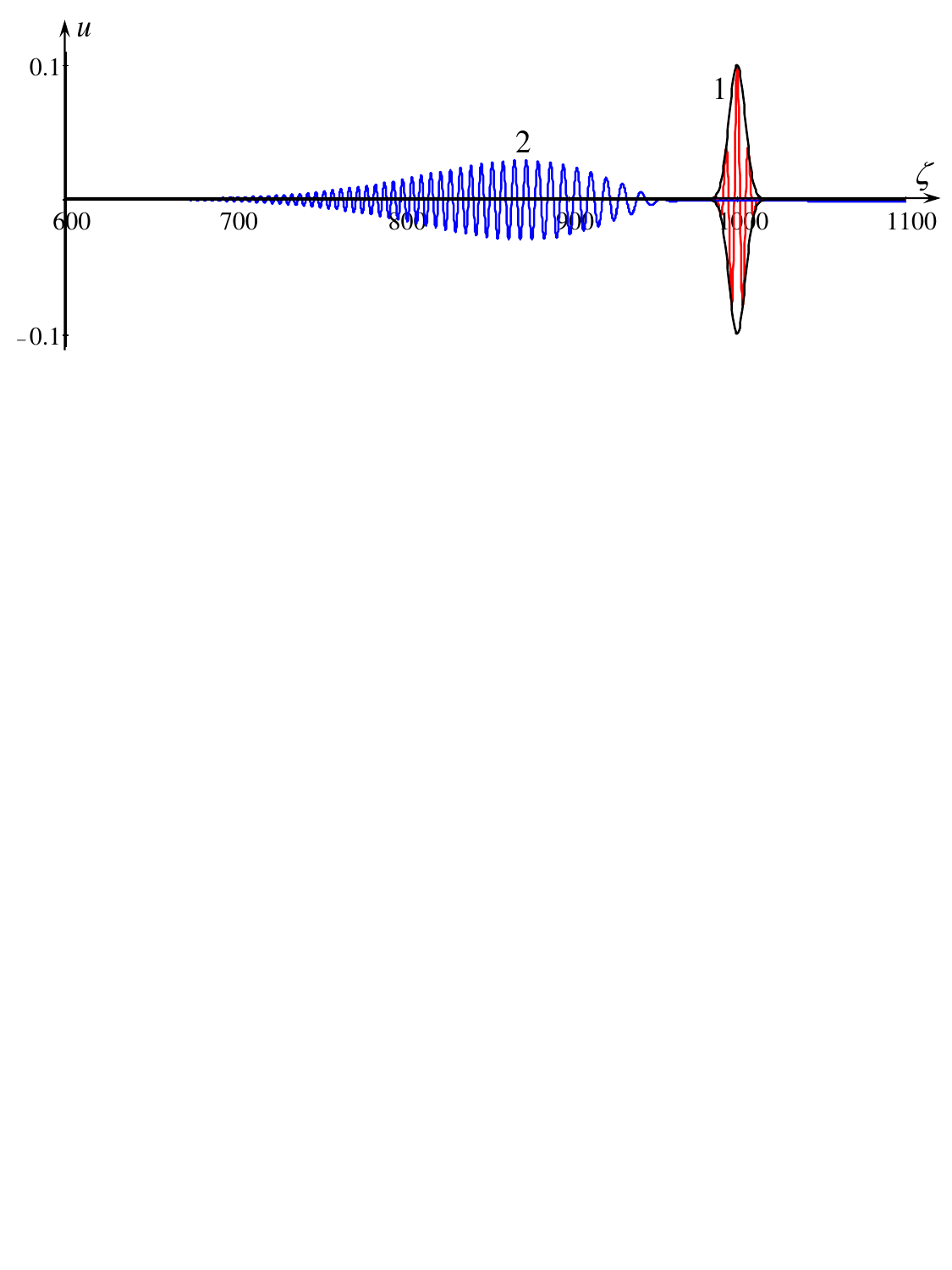}}} %
	\vspace{-14.0cm}%
	\caption{\small (Color online) The initial Gaussian pulse shown together with its envelope (line 1) and the result of its evolution at $\tau = 50$ (quasi-sinusoidal wave train 2).} %
	\label{f04}
\end{figure}

\begin{figure}[h!]
	{\centerline{\includegraphics [scale=0.75]{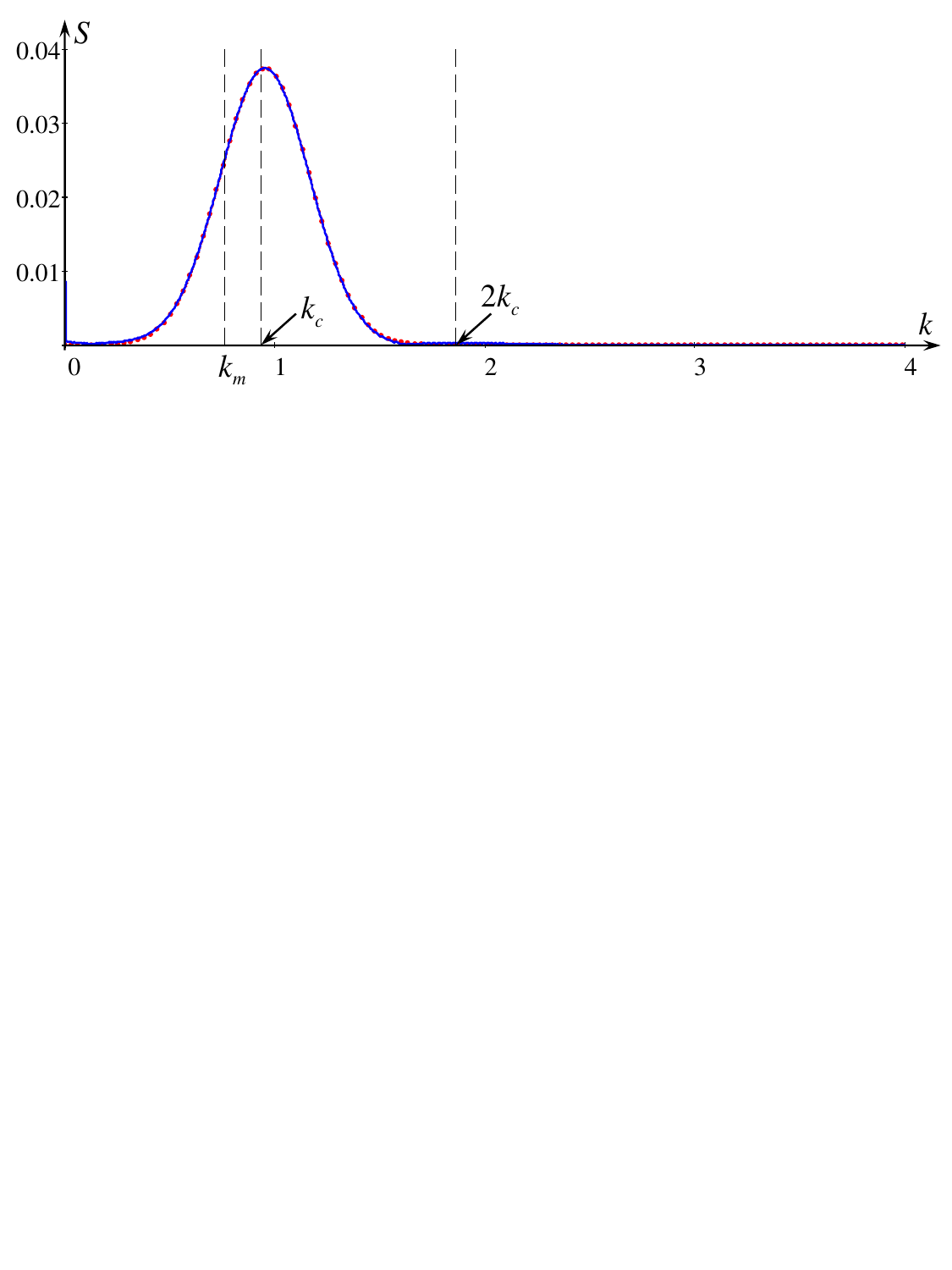}}} %
	\vspace{-14.0cm}%
	\caption{\small (Color online) The amplitude Fourier spectrum of the initial wave train with the Gaussian envelope shown in Fig. \ref{f04} (dotted line) and the spectrum of the wave train at $\tau = 50$ (solid line).} %
	\label{f05}
\end{figure}

A very similar situation occurred when we set the initial condition in the form of a Gaussian pulse with the carrier wave number $k_c > k_m$:
\begin{equation}
u(\zeta, 0) = A_0\exp{\left[-\left(\frac{\zeta - \zeta_0}{D}\right)^2\right]}\cos{k_c \zeta},
\label{Gaussian}
\end{equation}
where $A_0 = 0.1$, $D = 2\pi/k_c$, and the dimensionless wave number $k_c = (3/4)^{1/4}$. The solution was obtained on the interval $0 \le \zeta \le L$, where $L = 2000$. The spectrum of such initial perturbation is fairly wide, so that its significant portion is notably below the cut-off wave number $k_m$. Therefore such wave train fairly quickly decays and generates wave quasi-sinusoidal components with very small wave numbers. In Fig. \ref{f04} one can see the initial pulse together with its envelope (line 1) and the result of its evolution at $\tau = 50$ (quasi-sinusoidal wave train 2).
The Fourier spectra of both these signals are shown in Fig. \ref{f05}. It is clearly seen that the significant portion of the spectra are in the wave number range below the cut-off wave number $k_m$.

\begin{figure}[h!]
	{\centerline{\includegraphics [scale=0.75]{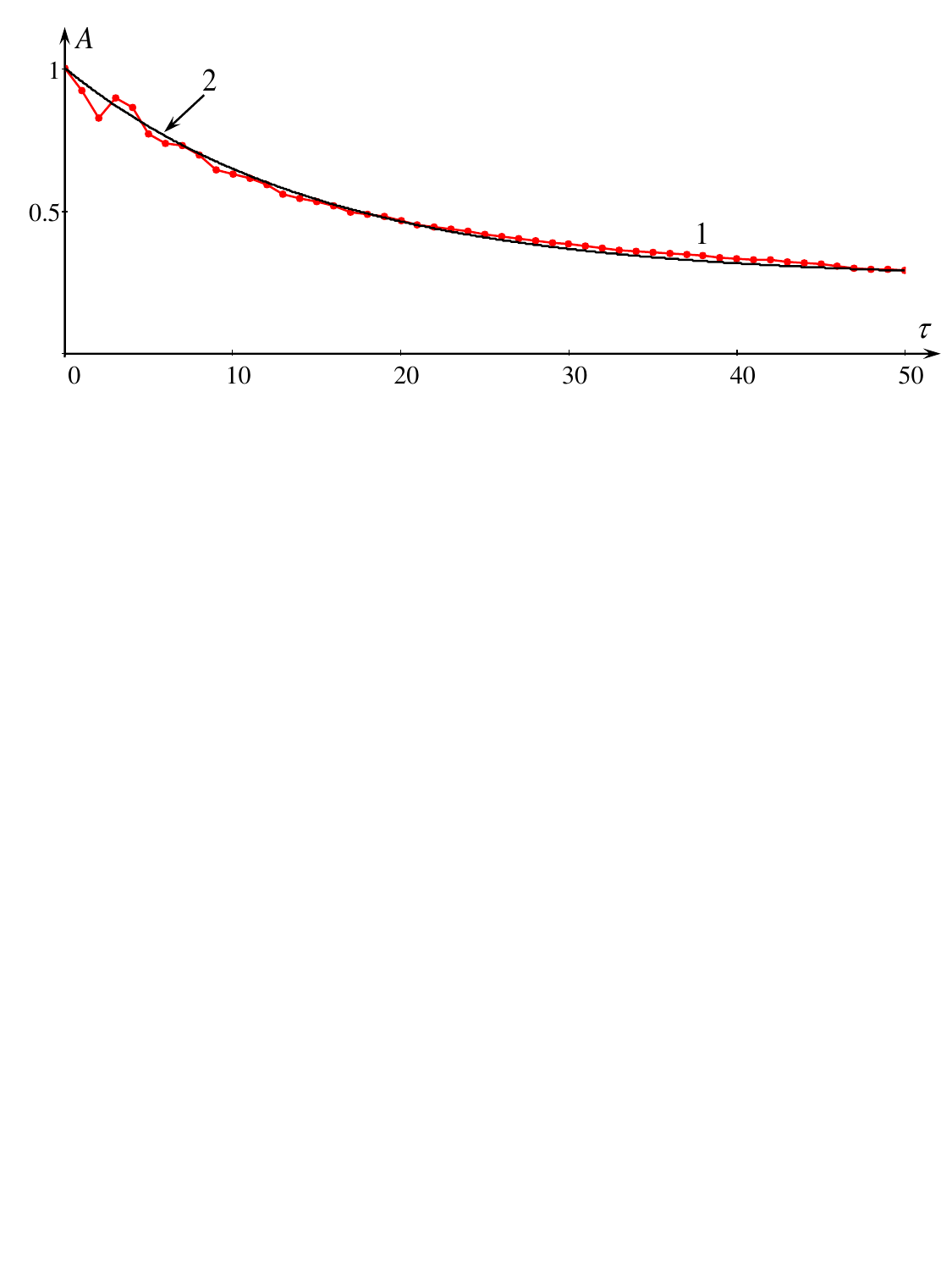}}} %
	\vspace{-13.5cm}%
	\caption{\small (Color online) The dependence of wave train amplitude $A(\tau)$ on time in the process of its decay (line 1). Line 2 shows the best fit approximation $A(\tau) = 0.74e^{-0.065\tau} + 0.26$.} %	
	\label{f06}
\end{figure}

In the process of evolution the initial Gaussian pulse gradually disperse, and its amplitude decreases (see Fig. \ref{f06}). Due to the nonlinearity one can observe generation of second harmonics and sub-harmonics with very small wave numbers in the vicinity of $k = 0$.

\begin{figure}[h!]
	{\centerline{\includegraphics [scale=0.75]{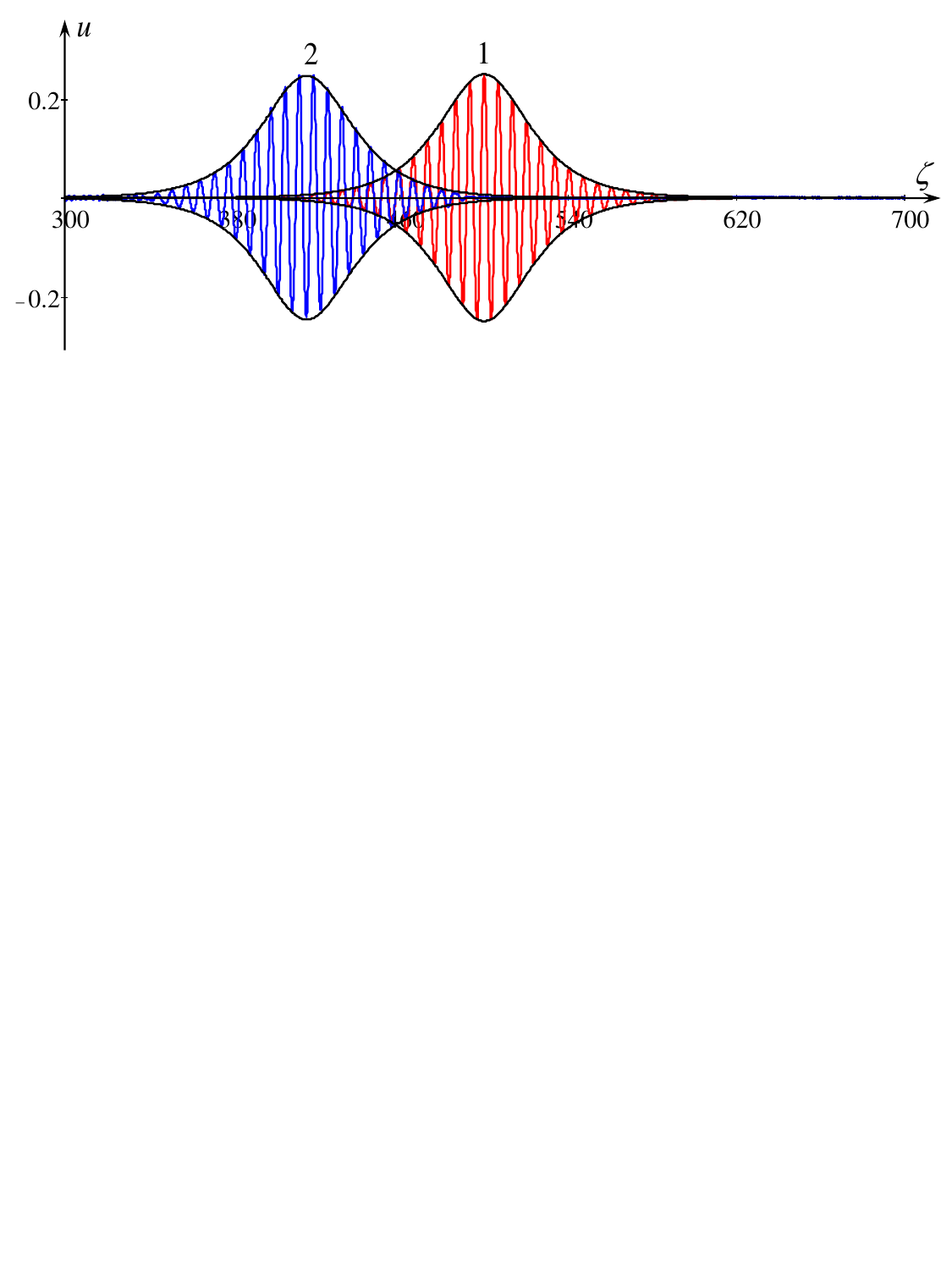}}} %
	\vspace{-13.5cm}%
	\caption{\small (Color online) The NLS soliton (\ref{EnvNLSSol}) of amplitude $A_0 = 0.25$, wave number $k = k_c$, and gauge $\kappa = 0$ shown together with its envelope (line 1) and the result of its evolution at $\tau = 25$ (line 2).} %	
	\label{f07}
\end{figure}

\begin{figure}[h!]
	{\centerline{\includegraphics [scale=0.75]{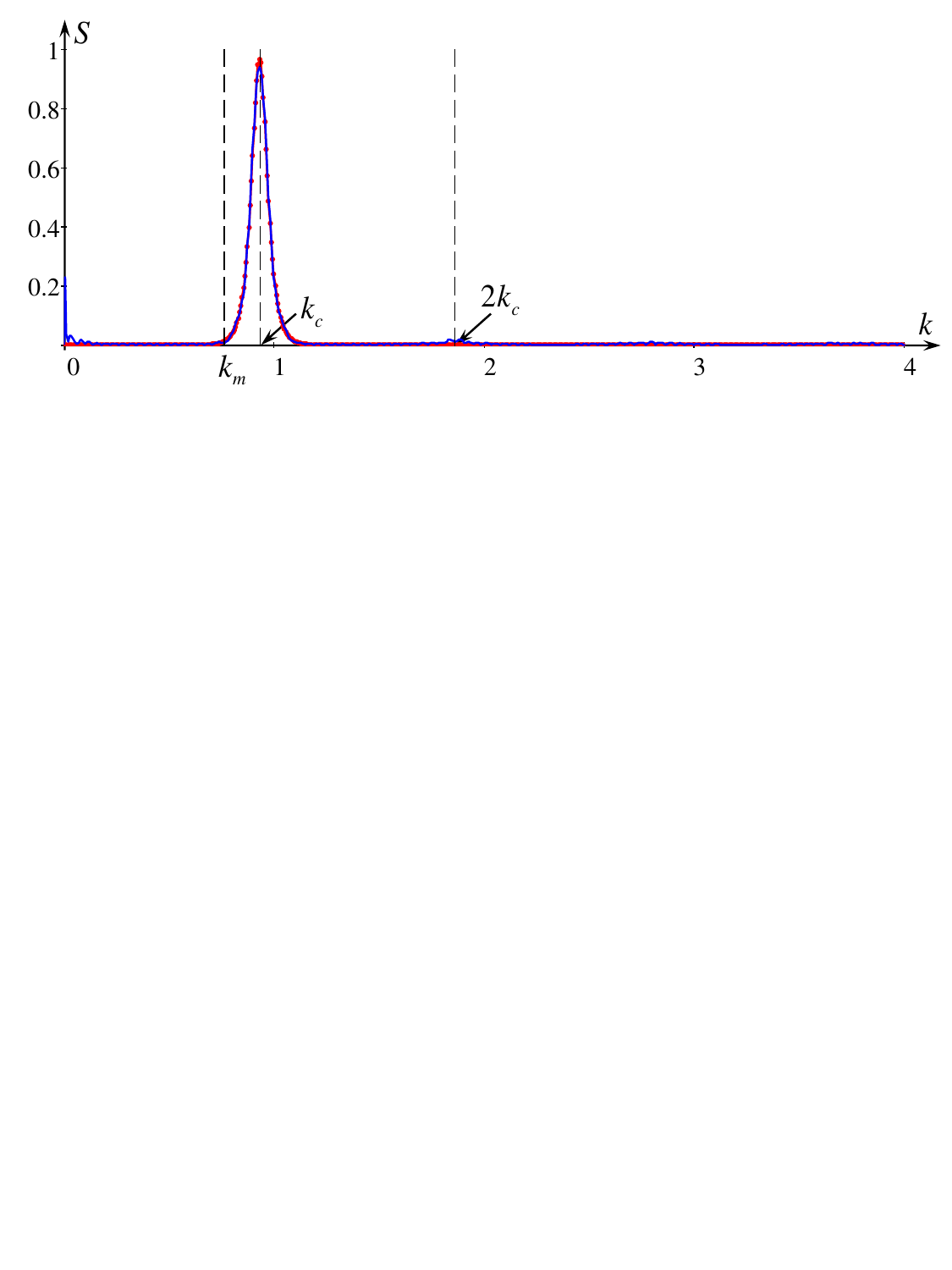}}} %
	\vspace{-13.5cm}%
	\caption{\small (Color online) The amplitude Fourier spectrum of the initial NLS soliton (dotted line) and the spectrum of the wave train at $\tau = 25$ (solid line).} %
	\label{f08}
\end{figure}

If the initial pulse is the NLS soliton (\ref{EnvNLSSol}) with the small amplitude and narrow spectrum, then it remains stable and propagates with the stationary envelope in accordance with the theoretical prediction. An example is shown in Fig. \ref{f07}, where the initial condition (line 1) was chosen in the form of NLS soliton (\ref{EnvNLSSol}) with the amplitude $A_0 = 0.25$, wave number $k = k_c$, and gauge $\kappa = 0$. As one can see, this soliton moves to the left with the negative group velocity as predicted.
The Fourier spectra of both these signals are shown in Fig. \ref{f08}; they are almost entirely above the cut-off wave number $k_m$, and only small insignificant portion is still below the cut-off wave number. There is still inevitable generation of second harmonic and sub-harmonic, but this does not crucially impacts on the soliton.

Thus, we can conclude that the dynamics of small-scale initial perturbations in the form of a wave train depends on the amplitude and spectrum of the perturbation. If the amplitude is small enough and spectrum is narrow so that it is entirely located above the cut-off wave number $k_m$, which separates the ranges of modulation stability $k < k_m$ and instability $k > k_m$, then the wave train can evolve into the envelope soliton which can be described either by the conventional NLS equation (\ref{StandardNLS}) or by one of its generalized versions (\ref{nls}). But if the spectrum of the initial wave train is wide and contains a portion which belongs to the range of modulation stability $k < k_m$, then such wave train either disperse completely or, possibly, in the result of evolution will produce a small-amplitude and narrow spectrum envelope soliton, whereas other portion of its initial energy will completely disappear due to dispersion.

In the conclusion to this section we present estimates of soliton parameters in the dimensional physical units. Considering typical parameters for the QGP in external magnetic fields of astrophysical objects \cite{fsn-18,fsn-19,we16}, we set the bag constant $70\,MeV/fm^{3}$, the chemical potential for all quarks $300 \, MeV$, the coupling constant $g=0.05$, the dynamical gluon mass $m_{G}=300\, MeV$, the magnetic field strength $B=10^{17}\,G $, the background density $\rho_{0}=2\rho_{N}$, and sound speed ${c_{s\,\perp}}\cong 0.2$ (for some details, see also the Appendix).  Choosing the dimensionless amplitude $A_{0}=0.05$ of a stable NLS soliton such as shown in Fig. \ref{f07}, we obtain for the amplitude of the baryon density perturbation of the $up$ quark: ${\delta\rho_{B}}_{u}=(\sqrt{\beta \Gamma}/\alpha)\times A_0 = 0.036$, where $(\sqrt{\beta \Gamma}/\alpha)=0.73$ (see Eqs. (\ref{ostrovsky}), (\ref{EnvNLSSol}) and (\ref{DLOstrEq})).  Then, for $B=10^{17}\,G$ the characteristic wavenumbers are  $k_{m}=2.42 \, fm^{-1}$, $k_{c}=2.96 \, fm^{-1}$, half-width of the envelope $\Delta = 22 \, fm$ and the group velocity of NLS soliton $V_{g}=-0.4$ .

\section{Conclusions}

In this paper we have demonstrated that weakly non-linear perturbations
in a quark-gluon plasma can be described by the Ostrovsky equation containing
two dispersion terms.  One of them  is responsible for the small-scale
dispersion (caused by the derivative terms ${\vec{\nabla^{2}}}{\rho_{B}}$
in Eqs. $(\ref{parap})$ and $(\ref{perp})$); the other one is responsible for the
large-scale dispersions (caused by the influence of the magnetic field).
The dispersion coefficients in the Ostrovsky equation are such that
stationary solitary wave solutions of this equation are impossible
\cite{Leonov,Galkin}. However when the large scale dispersion is
relatively small in comparison with the nonlinearity and small-scale
dispersion, then quasi-stationary solitary waves in the form of KdV
solitons can exist for a finite time experiencing a terminal decay
\cite{Grimshaw98}. After a long-term evolution, KdV solitons eventually
become envelope solitons which can be described by one of the
versions of the nonlinear Schr\"{o}dinger equation
\cite{rgs,Grimshaw08}. Originally it was suggested that the
NLS-type solitons are formed at the maximum of group speed, i.e.
at $k = k_m$ \cite{rgs}, however, later it was argued that
according to the analysis of numerical studies, such solitons are
formed at $k > k_m$ \cite{Whitfield}, although no suggestions were
proposed for the particular wave number of the carrier wave. In this
paper we have shown that the most likely NLS-type solitons will be
formed at $k = k_c > k_m$, where the growth rate of modulation
instability has a maximum (see also \cite{grimstepa}).

We have presented the results of the numerical study of a wave train
evolution with different carrier wave number within the framework
of the Ostrovsky equation. It has been demonstrated that the
perturbations with the narrow spectra of wave numbers can steadily
propagate in the form of envelope NLS-type solitons, whereas the
perturbations with relatively wide spectra gradually decay and disperse.

Thus, as follows from the analysis performed in this paper, envelope solitons in the form of Eq.(\ref{EnvNLSSol})
could exist in the quark stars as the microscopic perturbations; they could propagate over macroscopic distances carrying momentum and energy.
A big ensemble of such solitons can form a soliton gas with interesting properties (see, for example, Refs. \cite{redor,randoux}). The solitons contribution to the quark star
thermodynamics is an interesting issue which can be studied elsewhere.

\vspace{0.5cm}

\begin{acknowledgments}
This work was partially supported by the Brazilian funding agencies CAPES, CNPq and	FAPESP (contract 2012/98445-4). Y.S. acknowledges the funding of this study from the grant of President of the Russian Federation for state support of leading scientific schools of the Russian Federation (NSH-2685.2018.5). The authors are grateful to Prof. R. Grimshaw for the enlightening discussions.
\end{acknowledgments}

\appendix
\section{Derivation of the Ostrovsky equation}
%\label{ap:A}

As previously studied in \cite{fsn-19}, the background density upon which
small perturbations may occur, $\rho_{0}$, which
in this work is given by the typical QGP baryon density, is defined by multiples
of the ordinary nuclear matter density $\rho_{N}=0.17\, fm^{-3}$.

By using the reductive perturbation method (RPM) developed in \cite{rpm}, studied in
\cite{fsn-19,werev} and improved recently in \cite{azam,w2,javi17}, it is possible to preserve the nonlinear
terms such as the ${\vec{\nabla^{2}}}{\rho_{B}}$ terms in (\ref{parap})
and (\ref{perp}) to obtain nonlinear wave equations for the baryon density
perturbations.  We first use the pressure (\ref{parap}) and (\ref{perp}) to calculate the
gradient (\ref{gradepresses}) in Eq. (\ref{eulermag}).  This Euler equation with magnetic field effects included together with the EOS
and the continuity equation (\ref{contmag}) will be rewritten by the RPM approach, which consists in changing  variables going
from the $(x,y,z,t)$ space to the  $(X,Y,Z,T)$ space using the
``stretched coordinates'' defined by
$X=\sigma^{1/2}(x-{c_{s\,\perp}}\,t)\hspace{0.1cm}, \hspace{0.2cm}Y=
\sigma \, y\hspace{0.1cm}, \hspace{0.2cm} Z=\sigma \,z \hspace{0.1cm},
\hspace{0.2cm}  T=\sigma^{3/2}\,t$, and
$B=\sigma\, \tilde{B}$ for the magnetic field, where $\sigma \ll 1$ \cite{fsn-19}.  So, we
obtain the equations (\ref{eulermag}) and (\ref{contmag}) in the $(X,Y,Z,T)$
space containing the small
parameter $\sigma$, which is also the expansion parameter of the dimensionless
baryon density and dimensionless velocities \cite{fsn-19,rpm,werev}:
\begin{equation}
	\hat\rho_{B\,f}(x,y,z,t)={\frac{\rho_{B\,f}(x,y,z,t)}{\rho_{0}}}=1+\sigma
	{\rho_{f}}_{1}(x,y,z,t)+
	\sigma^{2} {\rho_{f}}_{2}(x,y,z,t) + \sigma^{3} {\rho_{f}}_{3}(x,y,z,t)+ \dots
	\label{roex}
\end{equation}
\begin{equation}
	{\hat{v}_{f\,x}}(x,y,z,t)={\frac{{v}_{f\,x}(x,y,z,t)}{{c_{s\,\perp}}}}=
	\sigma {{v_{{f}\,x}}_{1}}(x,y,z,t)+ \sigma^{2} {{v_{{f}\,x}}_{2}}(x,y,z,t) +
	\sigma^{3} {{v_{{f}\,x}}_{3}}(x,y,z,t)+ \dots
	\label{vxfex}
\end{equation}
\begin{equation}
	{\hat{v}_{f\,y}}(x,y,z,t)={\frac{{v}_{f\,y}(x,y,z,t)}{{c_{s\,\perp}}}}=
	\sigma^{3/2} {{v_{{f}\,y}}_{1}}(x,y,z,t)+ \sigma^{2} {{v_{{f}\,y}}_{2}}(x,y,z,t) +
	\sigma^{5/2} {{v_{{f}\,y}}_{3}}(x,y,z,t)+ \dots
	\label{vyfex}
\end{equation}
\begin{equation}
	{\hat{v}_{f\,z}}(x,y,z,t)={\frac{{v}_{f\,z}(x,y,z,t)}{{c_{s\,\parallel}}}}=
	\sigma^{3/2} {{v_{{f}\,z}}_{1}}(x,y,z,t)+ \sigma^{2} {{v_{{f}\,z}}_{2}}(x,y,z,t) +
	\sigma^{5/2} {{v_{{f}\,z}}_{3}}(x,y,z,t)+ \dots
	\label{vzfex}
\end{equation}

Equations (\ref{eulermag}) and (\ref{contmag}) in the $(X,Y,Z,T)$ space, rewritten in
terms of (\ref{roex}) to (\ref{vzfex}) are collected in
powers of $\sigma$, $\sigma^{3/2}$, and $\sigma^{2}$.  The terms for $\sigma^{n}$ with
$n > 2$ are neglected.  Each component of the Euler equation (\ref{eulermag}) in the RPM is given by:
\begin{equation}
	\sigma\Bigg\{-{\frac{\partial{{v_{f\,x}}_{1}}}{\partial X}}+
	\Bigg({\frac{9\,{g}^{2}\,{\rho_{0}}}{8\,m_{f}\,{m_{G}}^{2}\,
			({c_{s\,\perp}})^{2}}} \Bigg)
	{\frac{\partial{{\rho_{f}}_{1}}}{\partial X}}\Bigg\}
	+\sigma^{2}\Bigg\{-{\frac{\partial{{v_{{f}\,x}}_{2}}}{\partial X}}
	+{\frac{1}{({c_{s\,\perp}})}}{\frac{\partial{{v_{f\,x}}_{1}}}{\partial T}}
	+{v_{f\,x}}_{1}\,{\frac{\partial{{v_{f\,x}}_{1}}}{\partial X}}
	$$
	$$
	-{\rho_{{f}}}_{1}\,{\frac{\partial{{v_{f\,x}}_{1}}}{\partial X}}
	+\Bigg({\frac{9\,{g}^{2}\,{\rho_{0}}}{8\,m_{f}\,{m_{G}}^{2}\,
			({c_{s\,\perp}})^{2}}}\Bigg)
	{\rho_{{f}}}_{1}\,{\frac{\partial{{\rho_{f}}_{1}}}{\partial X}}
	+\Bigg({\frac{9\,{g}^{2}\,{\rho_{0}}}{8\,m_{f}\,{m_{G}}^{2}\,
			({c_{s\,\perp}})^{2}}}\Bigg){\frac{\partial{{\rho_{f}}_{2}}}{\partial X}}
	$$
	$$
	+\Bigg({\frac{9\,{g}^{2}\,{\rho_{0}}}{16\,m_{f}\,{m_{G}}^{4}\,
			({c_{s\,\perp}})^{2}}\
	}\Bigg){\frac{\partial^{3}{{\rho_{f}}_{1}}}{\partial X^{3}}}
	-{\frac{Q_{f}\,\tilde{B}}{m_{f}\,({c_{s\,\perp}})}}{v_{f\,y}}_{1}\Bigg\}=0,
	\label{eulerx}
\end{equation}
\\
\begin{equation}
	\sigma^{3/2}\Bigg\{-{\frac{\partial{{v_{f\,y}}_{1}}}{\partial X}}+
	\Bigg({\frac{9\,{g}^{2}\,{\rho_{0}}}{8\,m_{f}\,{m_{G}}^{2}\,
			({c_{s\,\perp}})^{2}}}\Bigg){\frac{\partial{{\rho_{f}}_{1}}}{\partial Y}}+
	{\frac{Q_{f}\,\tilde{B}}{m_{f}\,({c_{s\,\perp}})}}{v_{f\,x}}_{1}\Bigg\}
	+\sigma^{2}\Bigg\{-{\frac{\partial{{v_{f\,y}}_{2}}}{\partial X}}\Bigg\}=0,
	\label{eulery}
\end{equation}
and
\begin{equation}
	\sigma^{3/2}\Bigg\{-\Bigg({\frac{{c_{s\,\parallel}}}{{c_{s\,\perp}}}}\Bigg)
	{\frac{\partial{{v_{f\,z}}_{1}}}{\partial X}}+
	\Bigg({\frac{9\,{g}^{2}\,{\rho_{0}}}{8\,m_{f}\,{m_{G}}^{2}\,
			({c_{s\,\perp}})^{2}}}\Bigg){\frac{\partial{{\rho_{f}}_{1}}}{\partial Z}}\Bigg\}
	+\sigma^{2}\Bigg\{-\Bigg({\frac{{c_{s\,\parallel}}}{{c_{s\,\perp}}}}\Bigg)
	{\frac{\partial{{v_{f\,z}}_{2}}}{\partial X}}\Bigg\}=0.
	\label{eulerz}
\end{equation}
The continuity equation (\ref{contmag}) in the RPM is:
$$
\sigma\Bigg\{-{\frac{\partial{{\rho_{f}}_{1}}}{\partial X}}+
{\frac{\partial{{v_{f\,x}}_{1}}} {\partial X}}\Bigg\}+ \sigma^{2}
\Bigg\{-{\frac{\partial{{\rho_{f}}_{2}}}{\partial X}}+
{\frac{\partial{{v_{f\,x}}_{2}}} {\partial X}}+
{\frac{1}{({c_{s\,\perp}})}}{\frac{\partial{{\rho_{f}}_{1}}}{\partial T}} +
{\rho_{f}}_{1}{\frac{\partial{{v_{f\,x}}_{1}}}{\partial X}}+
{v_{f\,x}}_{1}{\frac{\partial{{\rho_{f}}_{1}}}{\partial X}}
$$
\begin{equation}
	+{\frac{\partial{{v_{f\,y}}_{1}}}{\partial Y}}+
	\Bigg({\frac{{c_{s\,\parallel}}}{{c_{s\,\
					\perp}}}}\Bigg)
	{\frac{\partial{{v_{f\,z}}_{1}}}{\partial Z}}\Bigg\}=0.
	\label{contmagos}
\end{equation}
Each bracketed factor multiplying the powers of $\sigma$ in the last four equations
must vanish  independently. So, the set of equations (\ref{eulerx}) to (\ref{contmagos}) when solved give the following equation:
$$
	{\frac{\partial}{\partial X}}\Bigg[{\frac{\partial{{\rho_{f}}_{1}}}{\partial T}}+
{\frac{3}{2}}({c_{s\,\perp}}){\rho_{f}}_{1}{\frac{\partial{{\rho_{f}}_{1}}}
	{\partial X}}
+\bigg( {\frac{{c_{s\,\perp}}}{4\,{m_{G}}^{2}}} \bigg){\frac{\partial^{3}
		{{\rho_{f}}_{1\
	}}}{\partial X^{3}}}\Bigg]
$$
\begin{equation}
	{} = {\frac{(Q_{f}\,\tilde{B})^{2}}{2{m_{f}}^{2}\,({c_{s\,\perp}})}}{\rho_{f}}_{1} - {\frac{({c_{s\,\perp}})}{2}}\Bigg(
	{\frac{\partial^{2}{{\rho_{f}}_{1}}}{\partial Y^{2}}}
	+{\frac{\partial^{2}{{\rho_{f}}_{1}}}{\partial Z^{2}}}
	\Bigg)
	\label{ostrostret}
\end{equation}
together with the constraint for the perpendicular speed of sound:
\begin{equation}
	({c_{s\,\perp}})^{2}= {\frac{9\,{g}^{2}\,{\rho_{0}}}{8\,m_{f}\,{m_{G}}^{2}}}
	\label{csperpcond}
\end{equation}
which is naturally found from the terms of $\mathcal{O}(\sigma)$. Equation
(\ref{csperpcond}) coincides with the ``effective sound speed'' obtained in the
linearization approach in our previous publications \cite{fsn-19,fsn-18}.
Neglecting the spatial third derivative term in the square brackets of Eq. (\ref{ostrostret}), we recover the results of paper
\cite{fsn-19}.

Equation (\ref{ostrostret}) in the original coordinates results in the
following nonlinear wave equation:
$$
	{\frac{\partial}{\partial x}}\Bigg[{\frac{\partial}{\partial t}}{\delta\rho_{B}}_{f}
+({c_{s\,\perp}}){\frac{\partial}{\partial x}}{\delta\rho_{B}}_{f}
+{\frac{3}{2}}({c_{s\,\perp}}){\delta\rho_{B}}_{f}{\frac{\partial}{\partial x}}
{\delta\rho_{B}}_{f}
+\bigg( {\frac{{c_{s\,\perp}}}{4\,{m_{G}}^{2}}} \bigg){\frac{\partial^{3}}
	{\partial x^{3}}}{\delta\rho_{B}}_{f}\Bigg]
$$
\begin{equation}
{} = {\frac{(Q_{f}\,B)^{2}}{2{m_{f}}^{2}\,({c_{s\,\perp}})}}{\delta\rho_{B}}_{f} - {\frac{({c_{s\,\perp}})}{2}}\Bigg(
{\frac{\partial^{2}}{\partial y^{2}}}{\delta\rho_{B}}_{f}
+{\frac{\partial^{2}}{\partial z^{2}}}{\delta\rho_{B}}_{f}
\Bigg).
	\label{ostro3d}
\end{equation}
Here we have used the identification ${\delta\rho_{B}}_{f}\equiv\sigma{\rho_{f}}_{1}$ in the
expansion (\ref{roex}), which is the first-order perturbation from the background
density $\rho_{0}$.  Equation (\ref{ostro3d}) is the Ostrovsky--Kadomtsev--Petviashvili equation describing wave propagation along the $x$-axis with a small diffraction in the perpendicular directions $y$ and $z$ \cite{PetvPokh,ostrov,OstrStep90,fnf-13}.
When the perturbation ${\delta\rho_{B}}_{f}$ does not depend on $y$ and $z$, then it reduces to purely Ostrovky equation (\ref{ostrovsky}).

\end{document}